\documentclass[10pt,twocolumn,letterpaper, table]{article}

\usepackage{iccv}
\usepackage{times}
\usepackage{epsfig}
\usepackage{graphicx}
\usepackage{amsmath}
\usepackage{amssymb}

\def\iccvPaperID{1235} 

\iccvfinalcopy 

\ificcvfinal
\def\assignedStartPage{9876} 
\fi

\ificcvfinal
\usepackage[breaklinks=true,bookmarks=false]{hyperref}
\else
\usepackage[pagebackref=true,breaklinks=true,colorlinks,bookmarks=false]{hyperref}
\fi

\ificcvfinal
\else
\fi

\usepackage{array}
\usepackage{xcolor}

\usepackage{multirow, makecell}
\usepackage{subcaption}
\usepackage[font=small]{caption}
\usepackage{booktabs}
\usepackage{amsmath}
\usepackage{nicematrix}
\usepackage{amsthm}
\usepackage{tikz}
\usetikzlibrary{matrix,calc}
\usepackage{pgfplots}
\pgfplotsset{compat=1.8}
\usepackage{amsfonts}
\usepackage[export]{adjustbox}
\usepackage{enumitem}
\usepackage{float}
\usepackage{bm}
\usepackage{dsfont}
\usepackage{bbold} 
\usepackage{arydshln}
\usepackage{hhline}

\DeclareMathOperator{\Tr}{Tr}

\newcommand{\bigO}[1]{\mathcal{O}(#1)}
\newcommand{\shorteq}{%
  \settowidth{\@tempdima}{-}
  \resizebox{\@tempdima}{\height}{=}%
}

\newcolumntype{L}[1]{>{\raggedright\let\newline\\\arraybackslash\hspace{0pt}}m{#1}}
\newcolumntype{C}[1]{>{\centering\let\newline\\\arraybackslash\hspace{0pt}}m{#1}}
\newcolumntype{R}[1]{>{\raggedleft\let\newline\\\arraybackslash\hspace{0pt}}m{#1}}
\newtheorem{problem}{Problem}[section]
\begin{document}

\title{Estimating Emotion Contagion on Social Media via Localized Diffusion in Dynamic Graphs}


\author{Trisha Mittal\\
University of Maryland\\
{\tt\small trisha@umd.edu}
\and
Puneet Mathur\\
University of Maryland\\
{\tt\small pmathur@umd.edu}
\and
Rohan Chandra\\
University of Texas, Austin\\
{\tt\small rchandra@utexas.edu}
\and
Apurva Bhatt\\
ShareChat, India\\
{\tt\small apurvabhatt@sharechat.co}
\and
Vikram Gupta\\
ShareChat, India\\
{\tt\small vikramgupta@sharechat.co}
\and
Debdoot Mukherjee\\
Meesho, India\\
{\tt\small debdoot@sharechat.co}
\and
Aniket Bera\\
University of Maryland\\
{\tt\small bera@umd.edu}
\and
Dinesh Manocha\\
University of Maryland\\
{\tt\small dmanocha@umd.edu}
}
\maketitle

\begin{abstract}
We present a computational approach for estimating emotion contagion on social media networks. Built on a foundation of psychology literature, our approach estimates the degree to which the perceivers' emotional states (positive or negative) start to match those of the expressors, based on the latter's content. We use a combination of deep learning and social network analysis to model emotion contagion as a diffusion process in dynamic social network graphs, taking into consideration key aspects like causality, homophily, and interference. We evaluate our approach on user behavior data obtained from a popular social media platform for sharing short videos. We analyze the behavior of $48$ users over a span of $8$ weeks~(over $200$k audio-visual short posts analyzed) and estimate how contagious the users with whom they engage with are on social media. As per the theory of diffusion, we account for the videos a user watches during this time (inflow) and the daily engagements; liking, sharing, downloading or creating new videos (outflow) to estimate contagion. To validate our approach and analysis, we obtain human feedback on these $48$ social media platform users with an online study by collecting responses of about $150$ participants. We report users who interact with more number of creators on the platform are $12\%$ less prone to contagion, and those who consume more content of `negative' sentiment are $23\%$ more prone to contagion. We will publicly release our code upon acceptance.
\end{abstract}



\section{Introduction}
Online social media platforms like Facebook, Twitter, and Reddit not only connect millions of people, but they also significantly impact society by sparking political discussions~\cite{metaxas2012social}, aiding disaster response~\cite{lazer2014parable,sakaki2010earthquake,merchant2011integrating}, and physically mobilizing people towards different causes~\cite{varol2014evolution,bond201261}. In contrast to these positive effects, a recent study led by Facebook~\cite{kramer2014experimental} highlighted one of the most subtle and least combated problems of digital content on social media--emotion contagion, which is defined as follows~\cite{goldenberg2019digital,ferrara2015measuring}:

\noindent \textit{\textbf{Emotion contagion} (EC) is a diffusion of emotions (positive or negative) and opinions over users in a social network such that the emotions and opinions of a ``perceiver'' become more similar to those of the ``expressor'' as a result of exposure to them.}
\begin{figure*}
  \centering
  \includegraphics[width=\textwidth]{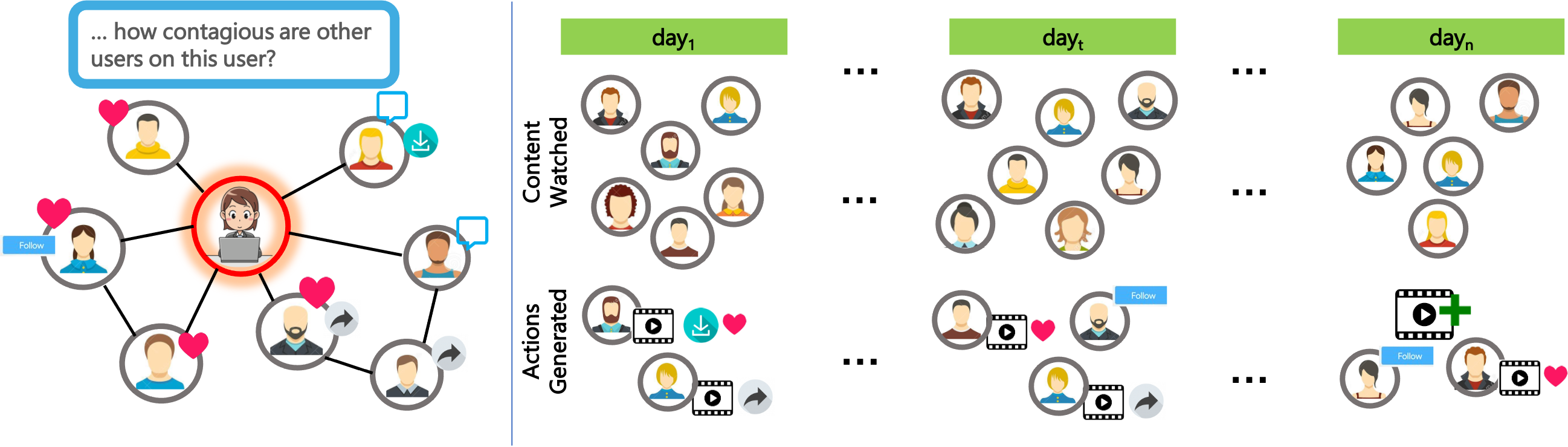}
  \caption{We propose a computational approach to estimate emotion contagion of digital content~(short video with audio posts) in online social networks. For instance, consider a user $c$, "perceiver"~(marked in red), on a social media platform engaging with $m$ users over a period of $t$ days. We analyze the \textit{inflow}, i.e., the content consumed by $c$, and the outflow, i.e., the actions that $c$ takes~(like, share, comment, follow, unfollow, download, create new content) in the same $t$ days. Our algorithm models the contagion phenomenon as a diffusion process and quantitatively estimates the degree to which the neighbor's ($1 \dots m$) emotions and content unintentionally match or influence.}
  \label{fig:teaser}
\end{figure*}
Emotion contagion can occur as a result of any type of exposure to the emotions of others. This can be broadly classified into non-digital~(face-to-face or telephonic) and digital~(social media) conversations. We now formally define Digital Emotion Contagion:

\noindent \textit{\textbf{Digital Emotion Contagion}~(DEC) is when EC occurs by sharing and expressing opinions on online platforms via multimodal digital content such as posts on Reddit and Facebook, tweets on Twitter, photos on Instagram, etc.}

While in both non-digital and digital emotion contagion, the emotions and opinions of ``perceivers'' change as a result of exposure to ``expressors'', the exposure is a lot more intense and frequent on digital media platforms (Figure~\ref{fig:teaser}) as all interactions on social media platforms are $1:n$ opposed to $1:1$ conversations in non-digital world. There are two main concerns with digital emotion contagion. First, users have little control over the content they consume on online social media platforms, putting them at risk of consuming harmful content~\cite{rideout2018digital,baruth2014psychological}. Second, social media platforms are known to incentivize emotion-rich content, leading to a self-reinforcing loop of enhanced emotion contagion~\cite{goldenberg2019digital}.


Social network analysis has previously focused on problems including hate-speech detection and filtering~\cite{mozafari2020hate}, violent content flagging~\cite{alvari2019detection} and viral post prediction~\cite{samuel2020message}. These problems have seen significant progress owing to the easy availability of huge datasets. However, prior work in emotion contagion research has been restricted to proving its existence on social media platforms~\cite{ferrara2015measuring,kramer2014experimental}, with very limited work on estimating emotion contagion. This is partly due to the absence of datasets capturing causal user behavior on social media, which prohibits collaborative research. This is primarily because this requires tracking user activity on social media platforms over a time span which is very sensitive data to publicly release. Prior works have also presented various hypotheses~\cite{lin2015emotional,he2016exploring,coviello2014detecting,bhullar2012self,gruzd2011happiness} about factors responsible for causing emotion contagion on social media. 

Furthermore, emotion contagion is not a widely understood term among social media users. Instead the closest concept that is well-understood is \textit{influence}. While influence is often intended, contagion occurs without the knowledge of the perceiver or expressor. Such similarities in shared emotions and opinions can be expressed on platforms using various mechanisms~(watching, liking, commenting, sharing, downloading a post, creating a new post, following and unfollowing other users, etc.). Emotion contagion is, at its core, a diffusion process, which can be characterized based on the following key aspects~\cite{ogburn2018challenges}:
\begin{enumerate}[noitemsep]
    \item Contagion is a \textbf{causal effect} on an ego's outcome at time $t$ of his alters outcome at time $s \leq t$.
    \item \textbf{Homophily} is ubiquitous and intransigent in the context of contagion, and it is the tendency of people who are similar to begin with to share network ties. 
    \item In a dyadic conversation, the contagion effect is well-defined, however social networks represent a paradigmatic opportunity for \textbf{interference}, where more than one subject could be responsible for subject's outcome.
\end{enumerate}

Our work focuses on estimating emotion contagion, a fairly nascent, but important line of research in social network analysis. Several prominent studies~\cite{goldenberg2019digital,ferrara2015measuring} have indicated that the focus in emotion contagion research should be to estimate contagion.

\noindent\textbf{Main Contributions:} We present the first computational approach for estimating emotion contagion in dynamic social network graphs. The input to our approach consists of a graph, $\bm{\mathcal{G}}$, where each node represents a user $i$ with profile $p_i$ and each edge between users $i,j$ represents the tie strength between $i$ and $j$. The objective of the approach is to compute the emotion contagion value, $\bm{\xi}$, for each node in $\bm{\mathcal{G}}$. The novel aspects of our work include:
\begin{enumerate}[noitemsep]
    \item We estimate emotion contagion by computationally modeling the key drivers of emotion contagion: causality, homophily, and interference. Despite the availability of many approaches that detect these factors, no method was previously known that could quantify them.
    
    \item Our diffusion approach models dynamic graphs; put simply, edges are bi-directional with different weights for each direction. Prior work on detecting emotion contagion operate on static uni-directional graph networks.
    
\end{enumerate}

We analyzed $48$ users' activity over a span of $8$ weeks and estimated the emotion contagion on them. We obtained the user behavior data from a popular social media video sharing platform. To validate our analysis, we obtain $150$ human feedback responses via user studies conducted based on the activity of these $48$ users. The user study corroborates our approach's results and quantifies homophily, causality, and interference.

\section{Related Work}
Section~\ref{subsec:online-social-media} highlights some of the recent work on analyzing social media content for various applications. In Section~\ref{subsec:emotion-contagion-theory}, we go over the theory of emotion contagion. In Section~\ref{subsec:digital-emotion-contagion}, we specifically focus on digital emotion contagion and discuss the challenges of existing research directions on emotion contagion in social media. 
Lastly, Section~\ref{subsec:diffusion-models} elaborates on prior literature in opinion propagation on social networks. 
\subsection{Analyzing Social Media Content}
\label{subsec:online-social-media}
The past two decades have witnessed an increase in the number of social media platforms, encouraging millions of users across the globe. The amount of content being generated and shared on these platforms is enormous and has given rise to many interesting research problems. One such direction is in automated systems for moderating content like hate speech~\cite{hate1, hate2}, violent content~\cite{violent} and fake news~\cite{fake1, fake2} on social platforms. Such platforms have also shown how useful they can be in response to disaster assessment~\cite{disaster1} and management~\cite{disaster2}. Other interesting research problems analyze content shared on these platforms to understand the dynamics of content likeability and social validation for content creators~\cite{like1}, influence and opinion propagation for social media marketing~\cite{opinion1,opinion2}, and the components that can make content trend and go viral on social media~\cite{samuel2020message,frame2018viral}. In this work, we analyze another such aspect, \textit{emotion contagion}, on social media platforms. 
\subsection{Theory of Emotion Contagion}
\label{subsec:emotion-contagion-theory}
Prior works have suggested that humans instinctively tend to align with the emotional states they perceive around them~\cite{ekman1983autonomic,hatfield1992primitive,barger2006service}. Various studies have concluded that emotions can be contagious~\cite{schachter1962psychological}, as a response to which individuals show behavioral, attentional, and emotional synchrony~\cite{hatfield1992primitive}. Prior literature has also associated emotion contagion to feelings of empathy and sympathy~\cite{hatfield1992primitive, dellarocas2007exploring} and emotional arousal~\cite{mehrabian1974approach, russell2003core, mehrabian1980basic}. The study of emotional contagion has been the focus of various disciplines because different types of interactions, such as commercial transactions, team communication, and human–robot interactions, can transfer emotions~\cite{li2017associations,chen2019leaders,kuang2019universality,manera2013susceptibility,matsui2019designing}. Marketing research on emotional contagion has focused on understanding how positive or negative emotions converge in positive or negative consumer behavior~\cite{dellarocas2007exploring,kramer2014experimental,fox2018face,cowley2014consumers}. More recently, emotion contagion through social media has been of heightened interest because of the high engagement on these platforms.
\subsection{Digital Emotion Contagion}
\label{subsec:digital-emotion-contagion}
Most prior works~\cite{kramer2014experimental, ferrara2015measuring,fan2014anger,coviello2014detecting} have conducted controlled experiments on social media platforms and confirmed the presence of emotion contagion and its manipulative effects on individuals. Similarly, \cite{tromholt2016facebook} and \cite{hunt2018no} show that the content we consume on social media affects not only the emotions that we express on these platforms but also our general well-being. As discussed in prior literature~\cite{goldenberg2019digital}, contagion can occur due to three mechanisms: (i) mimicry, (ii) activation, and (iii) social appraisal. More specifically, digital media platforms are known to incentivize competition for attention and positive reinforcement in the forms of likes or shares~\cite{brady2020attentional,brady2020mad}, and expressing emotions is an extremely useful way to attract attention. As a result, such emotion-rich digital activities lead to self-reinforcing loops that enhance emotion contagion over time. \cite{saldias2019tweet} developed \textit{Tweet Moodifier}, a Google Chrome extension that enables Twitter users to filter and visually mark emotional content in their Twitter feed to make them aware of and reflect on the emotion-rich content being consumed.

\subsection{Diffusion Models for Social Media Analysis}
\label{subsec:diffusion-models}
Diffusion models have increasingly been used to investigate how information propagates among people for various problems in social media analysis. More specifically, some of the classical learning models for opinion propagation and diffusion are threshold models~\cite{granovetter1978threshold}, with more recent generalisations by \cite{kempe2003maximizing}, and the De Groot or Lehrer-Wagner model~\cite{degroot1974reaching,lehrer2012rational}. Diffusion can be mathematically defined as an optimization problem with single objective of optimizing the goal of spreading information and capturing the rate of information dispersion. There are many factors which may influence the effects of information diffusion across social networks. Studies pointed out that diffusion-related behaviors are mainly caused by social infectivity and homophily~\cite{9037362,10.1145/1401890.1401897}. Information flow using diffusion models on social media with respect to viral tweets~\cite{hoang2018predicting}, pandemic information~\cite{meng2018diffusion,dinh2020covid}, and fake news and misinformation~\cite{6637343} has been widely studied. 

\section{Background and Problem Statement}
\label{sec: background_and_problem_statement}
To further enhance the readability and understanding of the paper, we first formally define the problem statement in Section~\ref{subsec:problem_statement}. In Section~\ref{subsec:diffusion-process}, we include a brief background on the mathematical diffusion process on which we base our approach. We then elaborate more on the key aspects of the contagion phenomenon and the factors that have been known to cause stronger contagion on social media as we use this as a base to build our model in Section~\ref{subsec:factors_emotion_contagion}. 
\subsection{Problem formulation}
\label{subsec:problem_statement}
Users on this social media platform have regular access to videos where they may watch videos belonging to a set of topics, $\mathcal{S}$, or perform an action from an action set $\mathcal{A}$, or both. Currently, we observe the following actions: \textit{play, like, download, share, create, follow,} and \textit{unfollow}. Each user $i$ has a watch history, $\mathcal{W}_i \in \mathbb{R}^{t\times \lvert \mathcal{S}\rvert\times n_i}$, indicating the videos that $i$ has watched over the past $t$ days, and an action history, $\mathcal{U}_i\in \mathbb{R}^{t\times \lvert \mathcal{S}\rvert\times n^a}$, where $a \in \mathcal{A}$ denoting the actions that $i$ has performed during those $t$ days.
\begin{problem}
Given as input a user $c$~(``perceiver''), with corresponding watch history $\mathcal{W}_c$ and action history $\mathcal{U}_c$, in a time period of $t$ days, we want to estimate the emotion contagion on $c$, denoted as $\bm{\xi}_c$, caused by its neighbors~(``expressors'').
\label{prob: main}
\end{problem}

\noindent We model emotion contagion among users $c$ via diffusion in a graph $\bm{\mathcal{G}} = \left (  \mathcal{V}, \mathcal{E}  \right )$. Each edge between $c$ and a user $1\leq i \leq m$ is bi-directional and weighted. We describe this diffusion process in the following section.

Emotion contagion can be modeled as a diffusion process over the social graph network. However, prior models fail to capture the computational aspects of emotion contagion in large networks due to the following two bottlenecks:



\begin{table}[t]
\centering
\resizebox{.9\columnwidth}{!}{
\begin{tabular}{ccc}
      \toprule
      \textbf{Aspect} & \textbf{Factors} & \textbf{References} \\ \midrule
      \multirow{2}{*}{Homophily} & Connection Strength  & \cite{lin2015emotional}\\
      & Age, gender, demographics & \cite{he2016exploring}\\
     \cmidrule{2-3}
     Causality & Time Gap b/w Content Consumed and Action Taken  & \\
     \cmidrule{2-3}
     Interference &  Sentiment      & \cite{coviello2014detecting, bhullar2012self, gruzd2011happiness}\\
      \bottomrule
    \end{tabular}
    }
    \caption{\textbf{Factors affecting emotion contagion:} We summarize factors suggested by prior literature that are known to cause stronger contagion. We model these factors in our approach (discussed in detail in Section~\ref{subsec:EC_factors})}
    \label{tab: ec_factors}
    \vspace{-10pt}
\end{table}


\begin{itemize}
    \item \textbf{Size of the graphs:} The cost of global diffusion to compute and store the histories and user profiles grows according to $\bigO{n^2}$ and $\bigO{n^3}$, respectively.
    \item \textbf{Static edge weights:} Static edge weights are easier to compute but do not accurately reflect the dynamic relationship between two users. Most prior works limit the edge weights to number of common friends, number of hops between connections, etc. Such information, though easily available, does not reflect the dynamic relationships between humans.
\end{itemize}
\begin{figure*}[t]
    \centering
    \includegraphics[width=.8\linewidth]{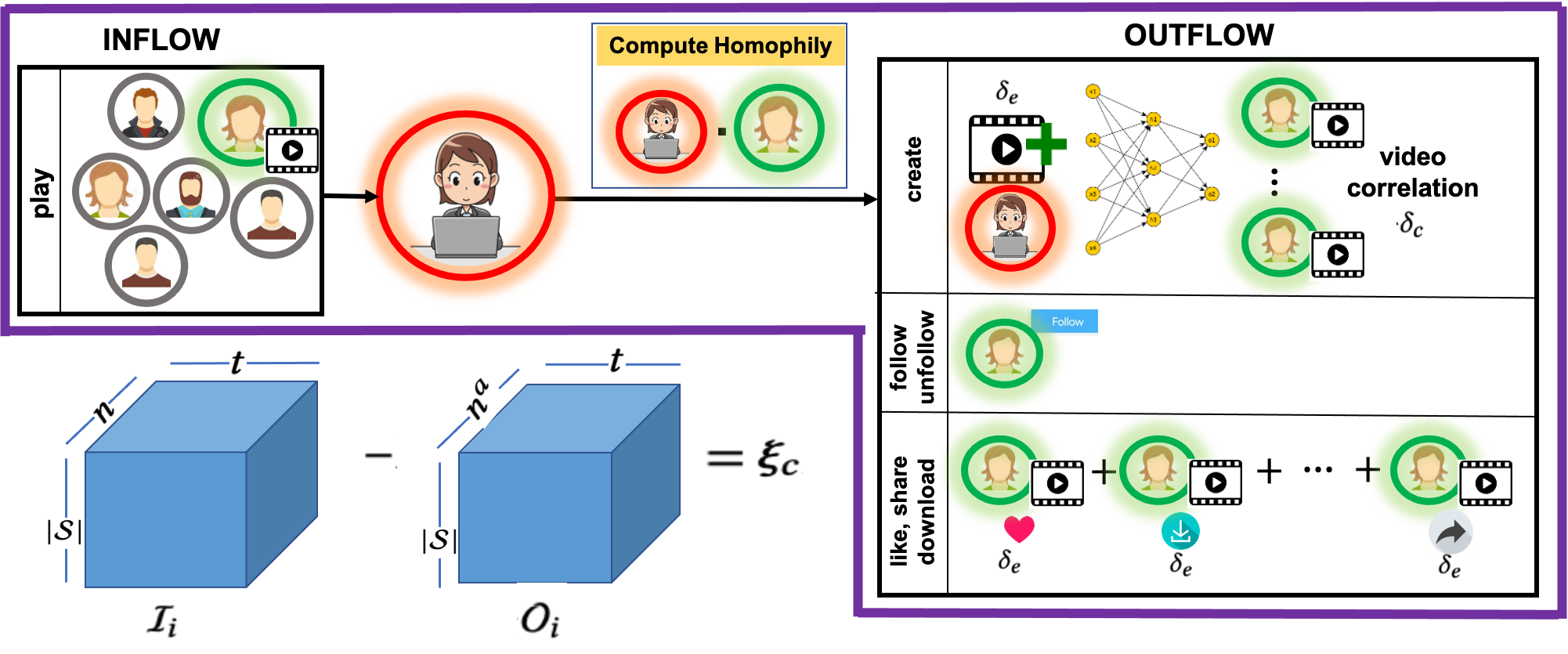}
    \caption{\textbf{Overview:} Given an input user $c$~(``perceiver'') and the set of $m$ neighbors of $c$, denoted as $\mathcal{M}$, we want to estimate the emotion contagion, $\bm{\xi}_c$, that  $\mathcal{M}$ causes $c$~(``expressors''). We begin by creating a graph of the social network, $\bm{\mathcal{G}_c}$ with $m+1$ nodes ($1$ central user node $c$ and $m$ nodes corresponding to $m$ neighbors). A dynamic process is allowed to occur where any neighbor $i, 1 \leq i \leq m$, may create content, and the central user $c$ consumes that content and may perform an action $a \in \mathcal{A}$ (part of Figure~\ref{fig: overview} outlined in blue). We proceed by using a combination of deep learning and social network analysis to model the various factors that characterize emotion contagion: \textit{homophily, causality, and interference}---and compute the inflow ($\mathcal{I}_i$) and outflow ($\mathcal{O}_i$) corresponding to any random neighbor $i$ (shown in green). Finally, we estimate the emotion contagion value by subtracting the outflow from the inflow (Section~\ref{subsec: main_model_EC}).}
  \label{fig: overview}
  \vspace{-10pt}
\end{figure*}

\subsection{Diffusion Process}
\label{subsec:diffusion-process}
We can model dynamic processes arising in information systems, such as traffic networks~\cite{nagatani2021traffic}, by performing diffusion on the associated graph structures described in the previous section. Let $\bm{\Phi}$ and $\bm{T}_{ij}$ represent the matter to be diffused and the velocity at which matter travels between nodes $i$ and $j$. The diffusion is described as 

\begin{equation}
        \frac{d\bm{\Phi}}{dt} = -  \bm{\widetilde L} \bm{\Phi} = -  \left(\bm{\widetilde D} - \bm{\widetilde A} \right) \bm{\Phi}
\label{eq: diffusion_eq}
\end{equation} 

\noindent Equation~\ref{eq: diffusion_eq} is the well-known diffusion equation~\cite{kondor2002diffusion}. We use $\bm{\widetilde L}$ to denote the weighted Laplacian representation of $\bm{\mathcal{G}}$. $\bm{\widetilde D}$ and $\bm{\widetilde A}$ represent the weighted degree and adjacency matrices of $\bm{\mathcal{G}}$, respectively. From Equation~\ref{eq: diffusion_eq}, it follows that the diffusion from a user $i$ to its neighbors $j$ is given as,

\begin{equation}
\begin{split}
        \frac{d\bm{\Phi}_i}{dt} &= -  \sum_j\left(\delta_{ij} \bm{\widetilde D}_{ii} -  \bm{\widetilde A}_{ij} \right) \bm{\Phi}_j\\
        &= -\left( \sum_j \bm{T}_{ij}\bm{\Phi}_i - \sum_j \bm{T}_{ij}\bm{\Phi}_j \right)= -\sum_j\bm{T}_{ij}\left(\bm{\Phi}_i - \bm{\Phi}_j \right),\\
        \end{split}
        \label{eq: in_out_background}
\end{equation} 

\noindent where $\bm{T}_{ij}$ denotes the edge weight. The second equality follows from the definition of the degree matrix and the fact that $\bm{\widetilde A} = \bm{T}$. $\sum_j\bm{T}_{ij}\bm{\Phi}_i$ and $\sum_j\bm{T}_{ij}\bm{\Phi}_j$ represent the outflow $\mathcal{O}_i$ ($j$ to $i$) and inflow $\mathcal{I}_i$ ($i$ to $j$), respectively. 
\subsection{Factors Affecting Emotion Contagion}

\label{subsec:factors_emotion_contagion}
Prominent studies~\cite{goldenberg2019digital, ferrara2015measuring} have propounded factors that can indicate strong or weak emotion contagion on online digital platforms. These factors are also summarized in Table~\ref{tab: ec_factors}. Various studies suggest that positive emotions are more prone to contagion than negative emotions~\cite{coviello2014detecting, gruzd2011happiness}. It has also been shown that stronger ties between the expressor and perceiver lead to stronger contagion~\cite{lin2015emotional}.
On the other hand, perceivers' personalities~\cite{cao2017method}~(easily influenced/agreeable), their online activities~\cite{del2016echo}, and their demographic features like age, gender, and culture~\cite{he2016exploring} have proven to influence the degree of emotion contagion online. 

\section{Our Approach}
\label{sec: predicting_emotion_contagion}

In this section, we present our algorithm for estimating emotion contagion in social networks. To begin, we give an overview of our approach in Section~\ref{subsec:overview}. We describe our approach as a diffusion model in Section~\ref{subsec: main_model_EC} and elaborate on how we take into account, \textit{homophily, environmental confounding, causality}, and \textit{interference} in Section~\ref{subsec:EC_factors}

\subsection{Overview}
\label{subsec:overview}

We describe our overall approach in Figure~\ref{fig: overview}. Given an input user $c$~(``perceiver'') and the set of $m$ neighbors of $c$, denoted as $\mathcal{M}$, we want to estimate the emotion contagion, $\bm{\xi}_c$, that  $\mathcal{M}$ causes $c$~(``expressors''). 

In our approach, we begin by creating a graph of the social network, $\bm{\mathcal{G}_c}$ with $m+1$ nodes ($1$ central user node $c$ and $m$ nodes corresponding to $m$ neighbors). A dynamic process is allowed to occur where any neighbor $i, 1 \leq i \leq m$, may create content and the central user $c$ consumes that content and may perform an action $a \in \mathcal{A}$. This is indicated by the part of Figure~\ref{fig: overview} outlined in blue. We proceed by using a combination of deep learning and social network analysis to model the various factors that characterize emotion contagion: \textit{homophily, causality, and interference}.

In the following sections, we describe our approach to modeling emotion contagion as a diffusion process that addresses two bottlenecks, global diffusion and static edge weights, enabling large-scale emotion contagion modeling in large social networks.




\subsection{EC in Localized Dynamic Graphs}
\label{subsec: main_model_EC}

\subsubsection{From Global to Local Graphs:}
Very little of the literature studying emotion contagion focuses on its estimation due to the challenges of modeling diffusion on large social graphs. The cost of global diffusion for computing and storing the histories and user profiles grows according to $\bigO{n^2}$ and $\bigO{n^3}$, respectively. Furthermore, global diffusion on these large graphs inevitably includes nodes with large degrees, for example, nodes corresponding to celebrities, athletes, and so on, which create bottlenecks. We instead perform localized diffusion, focusing on selected central user nodes. This selection may be targeted or, in the worst case, random. Due to the sparsity of bottleneck nodes, localized diffusion helps in bypassing most bottlenecks.

For each central user $c$, we extract a star graph $\bm{\mathcal{G}_c}$ consisting of $\lvert \mathcal{V}_c \rvert = m+1$ user nodes with $c$ as the central node and $\lvert \mathcal{E}_c \rvert = m$ edges.

\subsubsection{From Static to Dynamic Edge Weights}

We follow the speed-matching model used by Nagatani~\cite{nagatani2021traffic}. According to the model, flow is represented by the concentration of matter at the source multiplied by the velocity at the destination node. We set $\bm{T}_{ij} \gets u_{ij} \neq \bm{T}_{ji} \gets u_{iji}$. Then the dynamic equivalent form of Equation~(\ref{eq: in_out_background}) becomes,

\begin{equation}
\begin{split}
    \frac{\nabla \Phi_i}{\nabla t} = \mathcal{I}_i - \mathcal{O}_i= \sum_{in=1}^{k_{in}} \Phi_j u_{ji} - \sum_{out=1}^{k_{out}} \Phi_i u_{ij}\\
\end{split}
\end{equation}

\subsubsection{Modeling the Inflow ($\mathcal{I}_i$) and Outflow ($\mathcal{O}_i$)}

In this section, we use the concepts of localized diffusion on dynamic graphs to define equations for the inflow and outflow terms. The inflow $(i \xrightarrow{watch} c)$ describes the videos that are posted by $i$ and have been watched by $c$, who may then choose whether or not to perform any action. These actions are represented in the outflow $(c \xrightarrow{action} i)$. The inflow and outflow represent the change in density at a particular node, which corresponds to the rate of diffusion, or the strength of contagion. If, for a particular user $i$, we consider the watch history, $\mathcal{W}_i \in \mathbb{R}^{t\times \lvert \mathcal{S}\rvert\times n}$, till time $t$, and we further restrict each entry of the watch history to be the number of videos, $n$, then we can rewrite the watch history as the following 2D matrix,



\begin{equation}\mathcal{W}_i=\left[
\begin{array}{cccc}
 \bm{n}_{11} &\dots&\bm{n}_{t1} \\
 \vdots&\ddots&\vdots\\
 \bm{n}_{1\lvert \mathcal{\mathcal{S}} \rvert }&\dots&\bm{n}_{t\lvert \mathcal{\mathcal{S}} \rvert }\\
\end{array}\right]_{\lvert \mathcal{\mathcal{S}} \rvert \times t }
\label{eq: W}
\end{equation}

\noindent Using Equation~(\ref{eq: W}), the inflow corresponding to neighbor $i$ may be compactly written as

\begin{equation}
    \mathcal{I}_i = \mathcal{T}W_t,
    \label{eq: inflow}
\end{equation}

\noindent where $\mathcal{T}A$ is a trace function operator on a 2D matrix $A$, and computes $\Tr{\left(\sqrt{A}^\top\sqrt{A}\right)}$ (sum of all entries of $A$) with $\Tr{(\cdot)}$ is the matrix trace operator. Similarly, the outflow can be represented by the 2D matrix,

\begin{equation}
\mathcal{U}_i=\left[
\begin{array}{cccc}
\bm{\Gamma}(n_{11}) &\dots&\bm{\Gamma}(n_{t1}) \\
 \vdots&\ddots&\vdots\\
 \bm{\Gamma}(n_{1\lvert \mathcal{S} \rvert })&\dots&\bm{\Gamma}(n_{t\lvert \mathcal{S} \rvert })\\
\end{array}\right]_{\lvert \mathcal{\mathcal{S}} \rvert \times t }, \ \bm{\Gamma}(n) = \sum_{l=1}^{n}e^{-age}\delta_f\delta_c\delta_e
\label{eq: U}
\end{equation}

\noindent which implies

\begin{equation}
  \mathcal{O}_i = \left(p_i^\top p_c\right)\mathcal{T}\mathcal{U}_i
    \label{eq: outflow}
\end{equation}
\begin{figure*}[t]
    \centering
        \begin{adjustbox}{minipage=\linewidth,scale=0.95}
    \begin{subfigure}[h]{.22\textwidth}
        \includegraphics[width=.9\linewidth]{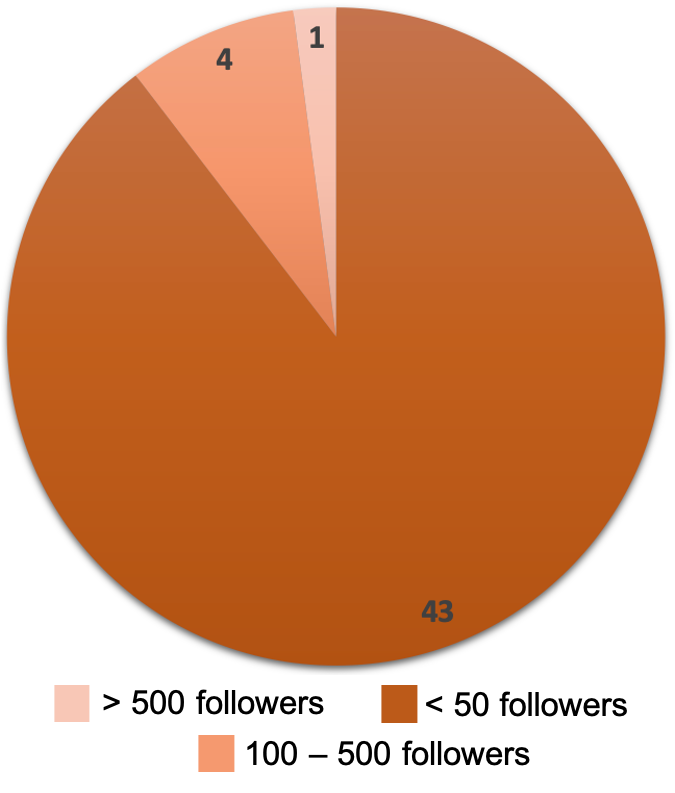}
        \caption{Depiction of the number of followers for the $48$ users analyzed.}
        \label{fig:followers-split}
    \end{subfigure}
    \quad
    \begin{subfigure}[h]{.22\textwidth}
        \includegraphics[width=.9\linewidth]{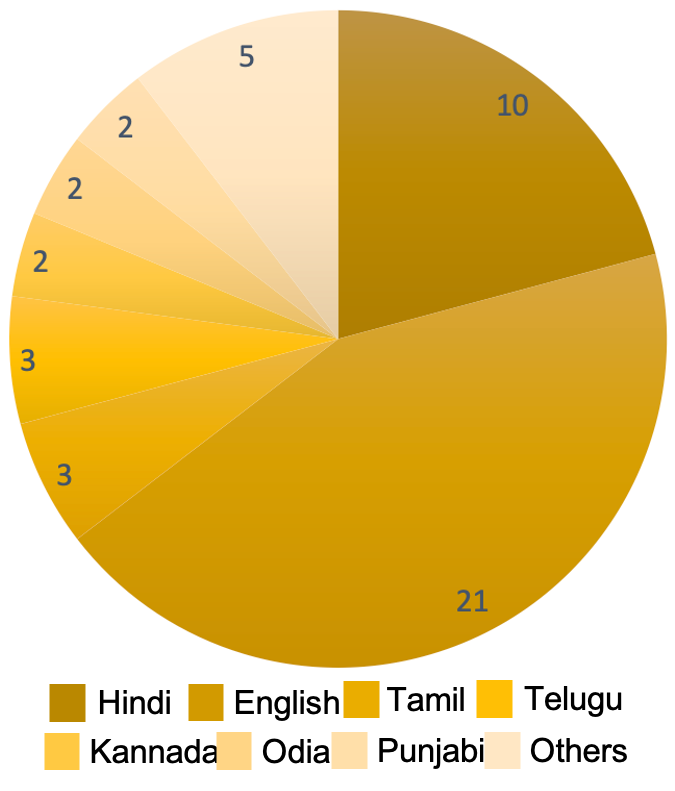}
        \caption{Depiction of the choice of language for the $48$ users analyzed.}
        \label{fig:language-split}
    \end{subfigure}
    \quad
    \begin{subfigure}[h]{.22\textwidth}
        \includegraphics[width=.9\linewidth]{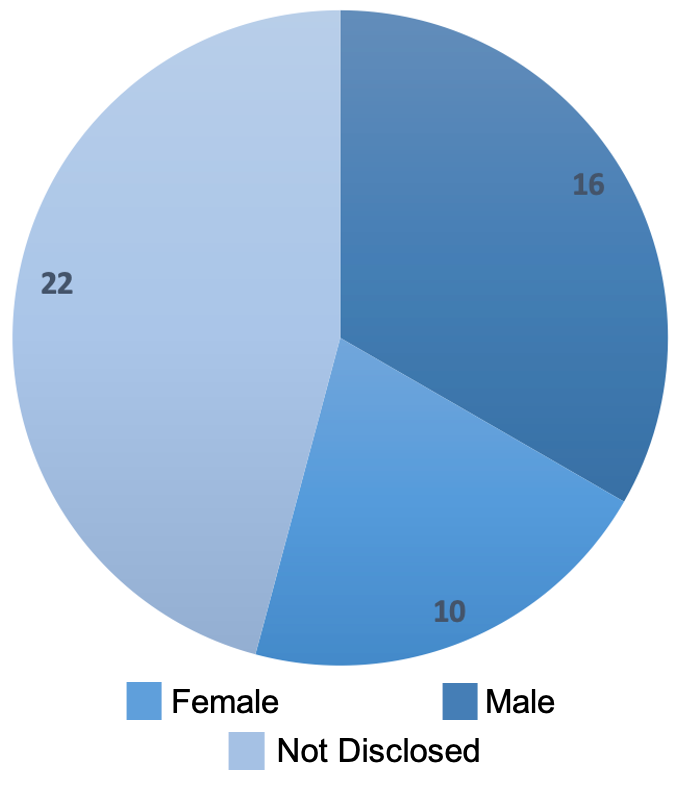}
        \caption{Depiction of the gender distribution for the $48$ users analyzed.}
        \label{fig:gender-split}
    \end{subfigure}
    \quad
    \begin{subfigure}[h]{.22\textwidth}
        \includegraphics[width=.9\linewidth]{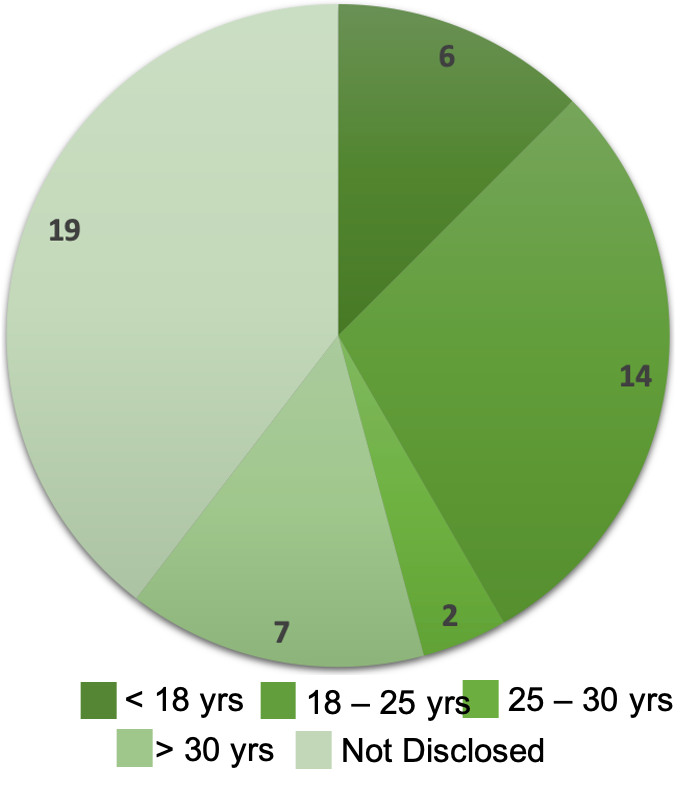}
        \caption{Depiction of the age distribution for the $48$ users analyzed.}
        \label{fig:age-split}
    \end{subfigure}
    \end{adjustbox}

    \quad
    \begin{subfigure}[h]{\textwidth}
        \centering
        \includegraphics[width=.8\linewidth]{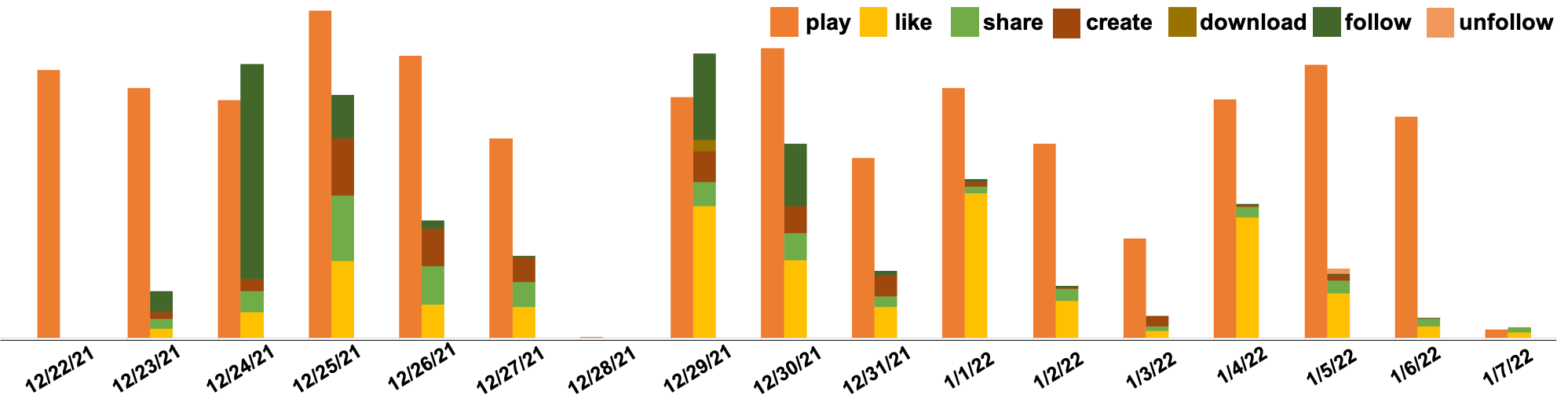}
        \caption{We visually depict the temporal aspect of the data. The visualization has been generated for one of the $48$ users for a period of $2$ weeks of the total $8$ weeks of data. For every day the user comes online, we have the videos played by the user~(orange column on the left) and also the data corresponding to like, share, download, create, follow and unfollow~(stacked column on the right).}
        \label{fig:temporal}
    \end{subfigure}
\caption{\textbf{Dataset Visualizations: }To further understand the data used for analysis, we depict user profiling statistics~(neighbors, age, gender, and language). We also show the temporality of the nature of the data used with an example of one of the $48$ users and their activities on the platform for $2$ weeks.  }
\vspace{-10pt}
\label{fig:dataset-vis}

\end{figure*}

We now define several key aspects that appear in the outflow equation (Equation~\ref{eq: U}). First, we decay the action taken by $c$ exponentially consider the temporal nature of the action. For example, if, on day $t$, $c$ \textit{likes} a video that was posted, in fact, on day $t-3$, then that action would be scaled by $e^{-3}$. Next, the actions \textit{follow, unfollow}, and \textit{create} do not contribute to the count directly and must therefore be modeled separately. $\delta_f$ is a step function with a value of $1$ if $c$ follows $i$ and $\frac{1}{2}$ otherwise. $\delta_c$ and $\delta_e$ denote the semantic and emotion correlation scores between a video created by $c$ and all other videos up to day $t$. For sentiment/emotion correlation, we use a context-based emotion prediction method~\cite{emoticon}. We use the audio, video and background context for inferring the sentiment. We obtain a single number $\pm 1$ referring to positive or negative sentiment videos.





\subsection{Modeling EC Factors}
\label{subsec:EC_factors}

Part of the difficulty in estimating emotion contagion is attributed to the challenges of modeling its underlying aspects: homophily, causality, and interference. These aspects, absent in related concepts like virality and influence, shape the contagion diffusion model. Estimating emotion contagion, therefore, is equivalent to modeling these factors.

\paragraph{Homophily:} Consider two sports fans, Tom and Harry. Both are young students who speak the same language, live in the same city, and follow each other on social media. Their connection is further strengthened due to the fact that both Tom and Harry react to each other's posts related to sports. This is an example of homophily with respect to personal demographics. We model this part of homophily by creating vector representations of users' information. Personal information, such as age, gender, language, city, connection strength to the central user, and so on, are first stored in a vector $\hat p$. We use $\hat p_c$ and $\hat p_i, 1 \leq i \leq m$ to denote the personal information vector of the central user and his or her neighbors, respectively. We then use multi-scale attributed node embedding~\cite{profile_embedding} to generate embeddings, $p_c, p_i$ from these raw information vectors. We then take the dot product, $p_c^T p_i$ to compute the correlation between the users.

Homophily not only considers the similarity between users, but also between the content they consume and post online. Consider, again, the two sports fans Tom and Harry, who regularly post news of soccer events. If Tom posts an announcement for an upcoming match, Harry is likely to like, share, or even save the details of the match. This is another example of homophily that points to the audio-visual and emotional signals of the content posted by Tom and consumed by Harry. Suppose, now, that Harry ends up attending that sports event and posts a video of (part of) the match. The audio-visual, as well as the sentiment, similarity of Tom's announcement and Harry's video is another indicator of homophily.

We measure audio-visual and sentiment similarity ($\delta_c and \delta_e$, respectively) between content using state-of-the-art deep learning models. Specifically, given a video $v_c$ created by a central user $c$, we compute it's audio-visual and semantic correlation, represented by $\delta_c$, with all the videos that $c$ has played. We use 3D convolutional networks~\cite{video_embedding} as a feature extractor function, denoted by $\mathcal{D}(\cdot)$, to compute the video embeddings and a python library~\cite{giannakopoulos2015pyaudioanalysis} to compute audio embeddings. We start by computing the correlation between the pair of vectors, $\mathcal{D}(v_c), \mathcal{D}(v_i)$ and $\mathcal{Q}(v_c), \mathcal{Q}(v_i)$,

\begin{equation}
    \bm{\rho}_{\mathcal{D}} = \frac {\operatorname {E} \big [\mathcal{D}(v_c)\mathcal{D}(v)\big]}{\sigma \big(\mathcal{D}(v_c)\big)\sigma\big(\mathcal{D}(v)\big)}, \quad
    \bm{\rho}_{\mathcal{Q}} = \frac {\operatorname {E} \big [\mathcal{Q}(v_c)\mathcal{Q}(v)\big]}{\sigma \big(\mathcal{Q}(v_c)\big)\sigma\big(\mathcal{Q}(v)\big)}
\end{equation}


\noindent for every video $v$ that $c$ has played. Finally, we manually apply a logarithmic scale to the correlation, according to our dataset. $\delta_c$ is computed as:

\begin{equation}
    \delta_c = -\textrm{age}\big(\log (1 - \bm{\rho}_{\mathcal{D}})\big) - \textrm{age}\big(\log (1 - \bm{\rho}_{\mathcal{Q}})\big)
    \label{eq: log_scaling}
\end{equation}

\noindent where the \textit{age} parameter indicates the causality factor, which is explained below. We note that Equation~\ref{eq: log_scaling} is a hand-crafted heuristic chosen after observing the given dataset. Learning this function using statistical learning and deep learning techniques is a promising future direction but we defer that to future work.

\paragraph{Causality:} Another factor used to estimate emotion contagion is the duration between when a perceiver plays some content created by the expressor and when he or she reacts to that content. Recall our case study from earlier; Harry, reacting to Tom's post, rather than a few days later, incurs different contagion levels; an immediate reaction is a case of higher contagion compared to the latter because of Harry's response time. This example perfectly illustrates the notion of causality: The faster a perceiver reacts to an expressor, the larger the causality and, by our model, the higher the contagion. More formally, we represent causality with the variable \textit{age} indicating the reaction response duration in days. Since contagion diffusion follows the standard linear decay model, the effect of \textit{age} decays exponentially. Hence, the term $e^{age}$ appears in Equation~\ref{eq: outflow}.

\paragraph{Interference:} Multiple neighbors may jointly influence the central user. While modeling single expressor-perceiver connections has been explored in prior studies, estimating contagion in the case of multiple perecivers and expressors is a harder endeavor. In our approach, we address this issue by leveraging the property of star traffic network graphs where the inflow and outflow are isolated along individual edges. To model the effect of $N$ neighbors, we simply sum the inflow and outflow along the $N$ distinct edges.

\section{Dataset}
\label{sec: dataset}
To the best of our knowledge, there is no dataset available for detection and estimation of emotion contagion, mainly because $i)$ this requires tracking a user's activity online for a consecutive time frame and $ii)$ contagion is not a well-understood term to be able to collect human annotations. Hence, for our purpose we extract users' social media information (content they watch, like, share, and download) from a popular video sharing platform designed for professional and amateur content creators. We extract user activity for $48$ users over a span of $8$ weeks on the platform. We give more insights about the data in Section~\ref{subsec: dataset-analysis}. Due to the sensitive nature of the extracted information, we have not released our dataset. That being said, to foster reproducability, we provide details for extracting our data in Section~\ref{subsec: dataset_extraction}.
\subsection{Dataset Structure and Extraction Process}
\label{subsec: dataset_extraction}
We select $48$ users who are active on the platform between November $27^\textrm{th}, 2021$ and February $1^\textrm{st}, 2022$. For each day a given user comes online, we extract the list of actions performed and their timestamps, URLs of the videos watched, created, shared, or downloaded, user IDs~(masked), and the topic of the videos. We also retrieve user profiling information of the $48$ users and their neighbors which includes age, gender, number of followers, demographic location, language. In total, this involved analyzing over $200$k short videos roughly watched, created, liked or shared by these users. 
\subsection{Dataset Analysis}
\label{subsec: dataset-analysis}
We visualize our data in Figure~\ref{fig:dataset-vis}. We present distributions over user profile information including followers, language, gender, and age in Figures~\ref{fig:language-split},~\ref{fig:gender-split}, and~\ref{fig:age-split}, respectively. In Figure~\ref{fig:temporal} shows the activity for $1$ central user during a $2$-week period. At a high level, we use the orange columns and multi-color stacked columns to calculate the inflow $\mathcal{I}$ and outflow $\mathcal{O}$, respectively. Visually, similar heights between an orange column and its corresponding stacked column indicate higher contagion ($\mathcal{I} - \mathcal{O} \approx 0$), which occurs on two occasions---$12/24/21$ and $12/29/21$. We provide a more in-depth analysis of the data in Appendix~\ref{sec:more-data-analysis}.

\section{Experiments and Results}
\label{sec:experiments}
We describe the user study conducted to obtain ground truth for the data, and analyze its responses, in Sections~~\ref{subsec:user-study} and~\ref{subsec:experiments-analysis}, respectively. We also discuss the efficiency of our approach in terms of computational resources highlighting the benefits of modeling localized emotion contagion in dynamic social network graphs in Appendix~\ref{subsec: efficiency_analysis}. We will publicly release our code upon acceptance.
\subsection{Obtaining Ground Truth via User Studies}
\label{subsec:user-study}
In the absence of benchmark datasets and curated ground truth for detecting and estimating emotion contagion, we conduct an extensive user study in the form of multiple questionnaires. Each questionnaire was designed to address three goals: $(i)$ to corroborate our approach's results using human feedback, $(ii)$ to understand people's interpretation of the importance over different actions, and $(iii)$ to emphasize the underlying aspects of emotion contagion, namely, homophily, causality, and interference. 

We prepared $10$ questions for each questionnaire. Questions $1-4$ (unique to each questionnaire) ask participants to answer questions about a given central user's online activity including his or her engagements with neighbors, question $5$ presents a scenario comparing different actions, and questions $6 - 10$ (identical across questionnaires) inquire about a participant's general social media usage. We circulate anonymous web links to these questionnaires and obtained approximately $150$ responses. In the following sections, we analyze the responses to $5$ of the questionnaires sent out. Throughout a questionnaire, we deliberately avoid the term contagion and, instead, use the term \textit{influence}. Due to lack of space, we attach full copies of these questionnaires in the supplementary material.
\subsection{Evaluation}
\label{subsec:experiments-analysis}
We divide our analysis based on the three parts described above. In all instances, a lower value for $\xi$ indicates higher contagion.
\subsubsection{\textbf{Analysis for Q$\bm{1-4}$: Does our approach estimate EC accurately ?}} 

We analyze the questionnaire as case studies. Our objective through these studies is to confirm that the contagion values obtained from our computational model agrees with the participants' responses.

\noindent \textbf{Case Study $\bm{1}$:} \textit{On a particular day, user A watched, and liked, $15$ videos posted by user B. On another day, user A watched $150$ videos posted by user D, without reacting to those videos. User A followed both users B and D afterwards.}

In this study, we asked participants to report which user, between B and D, had a greater influence on A in their opinion. Out of $17$ responses, $10$ indicated that B is likely to have had a bigger influence on A. This response strongly agrees with our computational model, which indicates that contagion caused by user B ($\xi_B = 11.73$) is approximately $11\times$ more than that caused by user D ($\xi_D = 124.56$). From the case study, we may also conclude that the ``active'' actions such as liking, sharing, etc. are stronger than ``passive'' actions like watching.

\noindent \textbf{Case Study $\bm{2}$:} \textit{User A watches, likes, and shares all videos created by Users B and C over a span of 8 weeks across various topics.}

As before, participants must report which user, from their perspective, had a greater influence on A. Unlike the previous one, however, this study does not contain user activity information. The participants do, however, have access to personal details about the users which includes age, gender, language, location, and number of followers. 

The response to this study was mixed. Of $28$ participants, $4$ chose B to have a greater influence, $6$ felt otherwise, and $18$ indicated there was not enough information to decide. Our model can explain why such a response was received. While, objectively, C does indeed have a stronger contagion effect than B ($\xi_C = 86.41$ versus $\xi_B = 133.50$), the profile embedding scores of B ($0.61$) and C ($0.65$) are similar. From the demographic information, C was closer to A in terms of age and language, but B was more popular with more followers, hence confusing the participants trying to decide who had a more similar profile to A.

\noindent \textbf{Case Study $\bm{3}$:} \textit{User A regularly watches videos of the topic, `Albums \& Concerts' created by various users on the platform for $4$ weeks, but does not create videos on the same topic. In the $5^\textrm{th}$ week, User A created the first $2$ videos of this topic.}

The question put to participants changes slightly in this study; instead of comparing the contagion tendencies between two users, we simply ask the participants if they felt A was influenced by the videos he or she watched during those four weeks. A majority ($20/28$) voted yes. To verify this, we compared the contagion value in the `Anger' topic of which A, similar to `Albums \& Concerts', watched many videos, but did not create any. We observed that when A did not create videos despite watching videos of that topic for four weeks, the contagion on A drops by a factor of $2$ ($\xi_\textrm{A\&C} = 39.00$ versus $\xi_\textrm{Anger} = 21.64$).
\subsubsection{\textbf{Analysis for Q$\bm{5}$: Do actions contribute equally to EC?}}
\begin{table}[H]
    \centering
    \resizebox{.9\columnwidth}{!}{
    \begin{tabular}{cccc}
    \toprule
         \textbf{Claim} & \textbf{User Study} & \textbf{Approach}\\
         \midrule
         follow/unfollow  $>$ like/share/download  & $73.2\%$ & $50\%$\\
         create $>$ follow/unfollow  & $57.1\%$ & $75\%$\\
         create $>$  like/share/download  &$64.7\%$ & $87.5\%$\\
         create $>$ watch  & $85\%$ & $98\%$\\
        \bottomrule
    \end{tabular}
    }
    \caption{\textbf{User Study (Q$5$) Results: }We summarize the results of Q$5$ and also report the performance of our approach on $48$ users with randomly selected creators.}
    \label{tab:5-analysis}
  \vspace{-10pt}
\end{table}

Question $5$ compares the potential for causing contagion between pairs of actions in $\mathcal{A}$. An example of such a question could be to compare \textit{liking} $10$ videos of a particular neighbor with \textit{creating} $1$ video in the same category. In this example, participants are asked to report which of the two actions, in their opinion, had a greater influence on the central user. We present $5$ such comparisons in Table~\ref{tab:5-analysis}. The first column contains the question setup containing the pairwise action comparison. The second and third columns indicate the percentage of participants and central users that agree with the corresponding relation in the first column. We refer readers to the questionnaires for exact language of the questions and options. 

This experiment further serves to distinguish influence from contagion. Actions that may seem influential over other actions may not necessarily cause stronger contagion. For instance, $73.20\%$ participants indicated that follow (or unfollow) induces a greater influence than actions such as like, share, and download, whereas our analysis yielded only $50\%$ of the central users who agreed with that assessment. Objectively, this may make sense since to follow someone is a ``stronger'' response. Our contagion model additionally takes into account profile similarity, age of the content, and causality to determine contagion. In the instance considered above, the central users may have followed users after considerable time passed since the former watched the latter's videos. Alternatively, the profiles of the central users and the expressors might have been dissimilar.
\begin{figure}[t]
    \centering
    \includegraphics[width=.95\columnwidth]{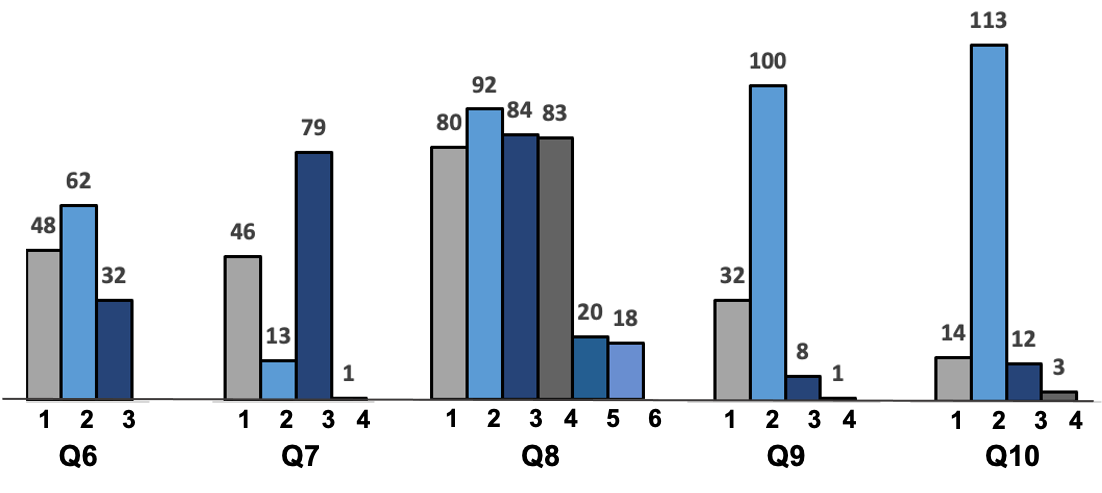}
    \caption{\textbf{User Study (Q6 - Q10) Results: }Response summary to the last five questions~(same across all the questionnaires).}
  \label{fig:6-10-analysis}
  \vspace{-10pt}
\end{figure}
\subsubsection{\textbf{Analysis for Q$\bm{6-10}$: Is there a discernible relationship between people's social media usage and the factors characterizing EC?}}

We asked the following questions:

    \textit{Q$6$: What do you think is the concept of ``digital emotion contagion''?}
    
    \textit{Q$7$: Do you think the content you share on social media is affected, influenced or based on the content you consume online? Additionally, do your opinions often change based on the content you consume on social media?}
    
    \textit{Q$8$: What comprises of the people you follow on social media?}
    
    \textit{Q$9$: How do you decide to follow/subscribe a user on social media?}
    
    \textit{Q$10$: When do you like/share a content post on social media?}

\noindent We summarize the results of the user study to questions $6-10$ in Figure~\ref{fig:6-10-analysis}. We refer the readers to the questionnaires in the appendix for the options corresponding to each question. 

Questions $6$ and $7$ survey participants' knowledge of digital emotion contagion and its effects on participants. From the responses to question $6$, $75\%$ of the participants are unaware of emotion contagion, misinterpreting contagion, instead, with virality (option $1$) or posts intended to influence their audience (option $2$). For question $7$, we found that that $57\%$ of the participants believe the content they share online is not influenced by what they consume while $42\%$ indicated otherwise. The lack of a clear consensus among the participants reveals that there is little awareness of the effects of contagion, emphasizing the importance of the proposed work.

\begin{table}[t]
    \centering
    \resizebox{.8\columnwidth}{!}{
    \begin{tabular}{cp{3cm}c}
    \toprule
     \textbf{S. No.}  & \textbf{Scenario} & \textbf{Approach Insights}\\
         \midrule
        $1$ & Watch more `negative'~(sentiment) than `positive' videos & $23\% \uparrow$\\
        $2$ & More neighbors & $12\% \downarrow$\\
        $3$ & More homophilic neighbors & $28\% \uparrow$\\
        $4$ & Topic diversity & $8\% \downarrow$\\
        \bottomrule
    \end{tabular}
    }
    \caption{\textbf{Other Scenarios: }We analyze some scenarios and generate results for $48$ users and understand how emotion contagion changes with different scenarios.}
    \label{tab:scenarios}
  \vspace{-10pt}
\end{table}

Our objective through questions $8, 9,$ and $10$ was to discover the presence of homophily, causality, and interference in emotion contagion. From the responses to question $8$, participants follow their friends~(option $2$), family~(option $3$), strangers whose content they relate to~(option $4$), and celebrities~(option $1$) which are not necessarily homophilic in nature. Our takeaway from Q$9$ is that actions on social media are causal in nature. $71\%$ of the users reported that they take their time before `following' people on social media. In Q$10$, $80\%$ participants reported that they base low-effort actions like `like' and `share' mostly on the content of the respective post~(option $2$), followed by liking because posts were created by close friends~(option $1$) or for bookmarking for inspiration of future posts~(option $3$), indicating each relationship on social media leads to a different level of engagement and hence will be contagious at varying levels. 
Responses to these questions validate the need and the decision of taking into consideration, aspects of homophilic connections, causality of actions, and the interference of contagious connections.
\subsection{Scenario Specific Insights}
Our approach offers flexibility to test, and generate, insights of contagion in a wide range of settings. We examine four such settings summarized in Table~\ref{tab:scenarios}. In the first experiment, we vary the sentiment of the content consumed by central users. We find that the contagion increases when the content is more negative than positive (Row $1$). In fact, such a finding was theoretically hypothesized by Goldenberg and Gross~\cite{goldenberg2019digital}. In the second experiment, we investigated the effects of increasing the number of \textit{homophilic} neighbors for each central user, and observed an increase in contagion~(Row $3$). Finally, in rows $2$ and $4$, we observe that contagion decreases when central users interact with more expressors, who may not necessarily be homophilic, or engage in more diverse topics, which decreases the exposure to content per topic.  

\section{Conclusion}
We shed light on a crucial, yet unexplored from a computational perspective, research problem that affects millions of users. For the first time, using a combination of deep learning and social network analysis, we showed we can accurately estimate emotion contagion in dynamic social media networks. We verified our approach by conducting a user study conducted with $150$ participants and comparing the participant responses to the outputs from our approach. 

There are some limitations to this work. Currently, we heuristically scale the outflow by the audio-visual and semantic similarity (Equation~\ref{eq: U}). This equation, fine-tuned to our dataset, may need to be adjusted for other datasets. Second, we do not consider the environment of a user off the social media platforms. The challenges of procuring users' environment information notwithstanding, such information contributes to the confounding aspect of emotion contagion, but is not considered in this paper, since it requires specialized data. In the future, we plan to explore ways of automatically estimating the outflow equation from the data using machine learning. We discuss ethical considerations taken with the handling of data in Appendix~\ref{sec:ethical-considerations}.

{\footnotesize
\bibliographystyle{plain}
\bibliography{refs}
}
\appendix
\clearpage
\section{More Data Analysis}
\label{sec:more-data-analysis}
\begin{figure*}
  \includegraphics[width=\linewidth]{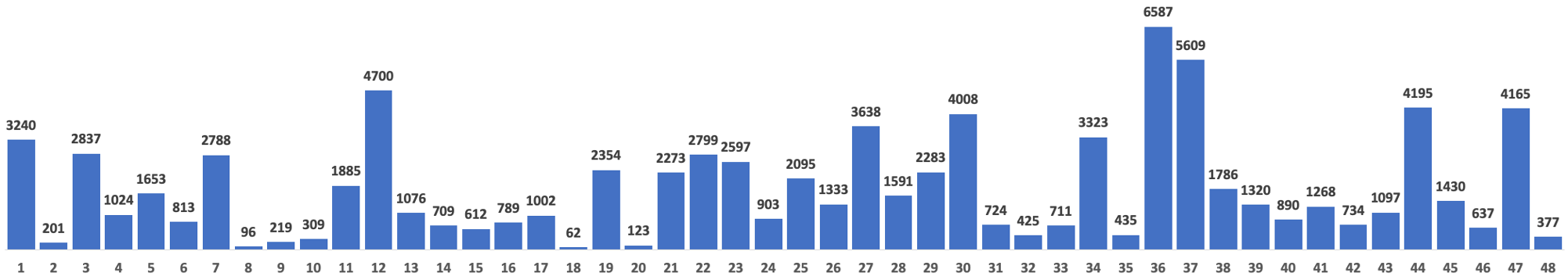}
  \caption{Depiction of the number of neighbors each of the $48$ users interacted with over $8$ weeks.}
  \label{fig:neighbors}
 \end{figure*}

\begin{figure*}
  \includegraphics[width=\linewidth]{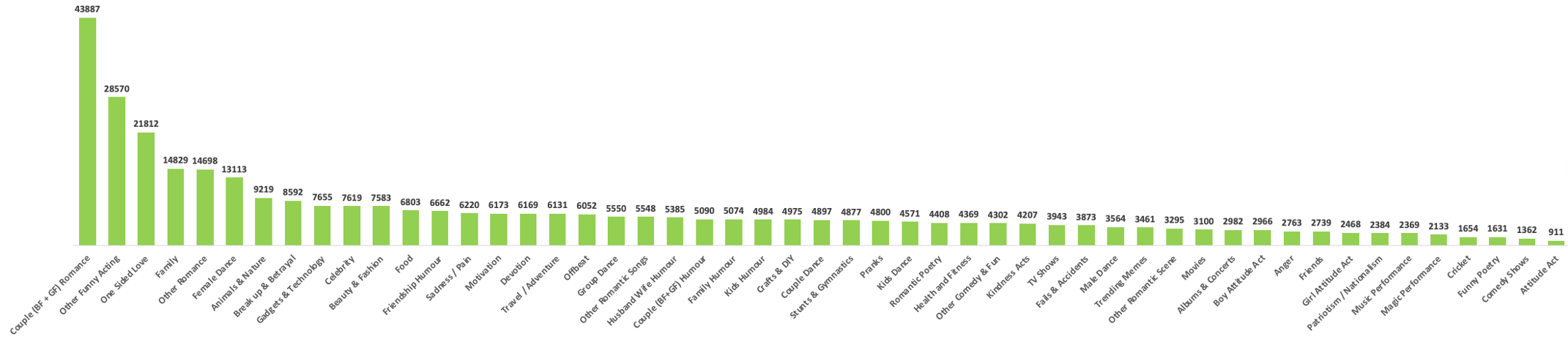}
  \caption{Depiction of the breadth of topics the $48$ users engage in  over $8$ weeks.}
  \label{fig:topic}
 \end{figure*}

\begin{figure*}
  \includegraphics[width=\linewidth]{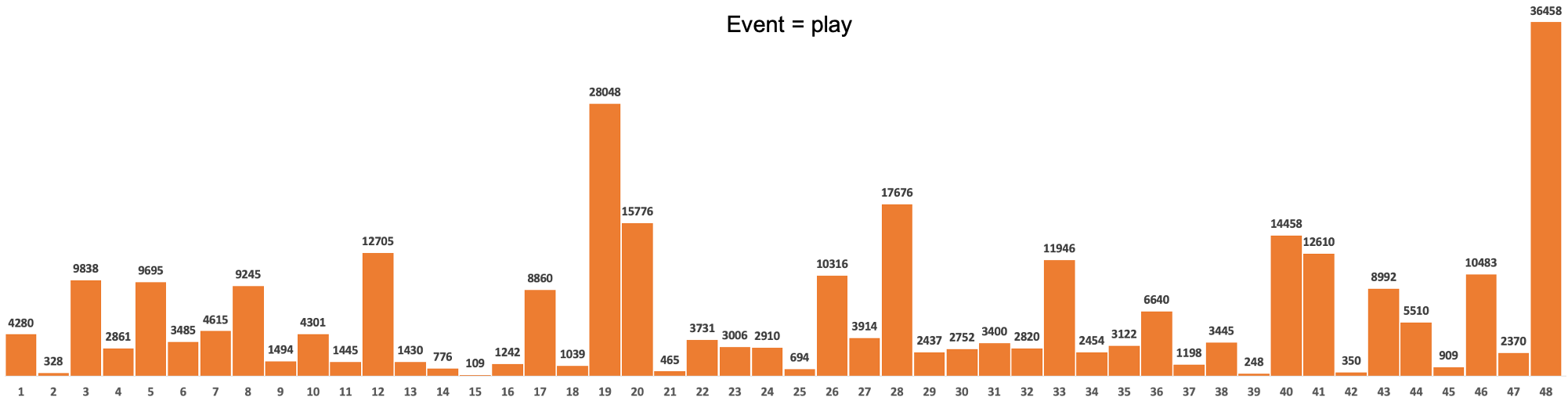}
  \caption{Depiction of the number of videos the $48$ users watched in  over $8$ weeks.}
  \label{fig:play}
 \end{figure*}

\begin{figure*}
    \includegraphics[width=\linewidth]{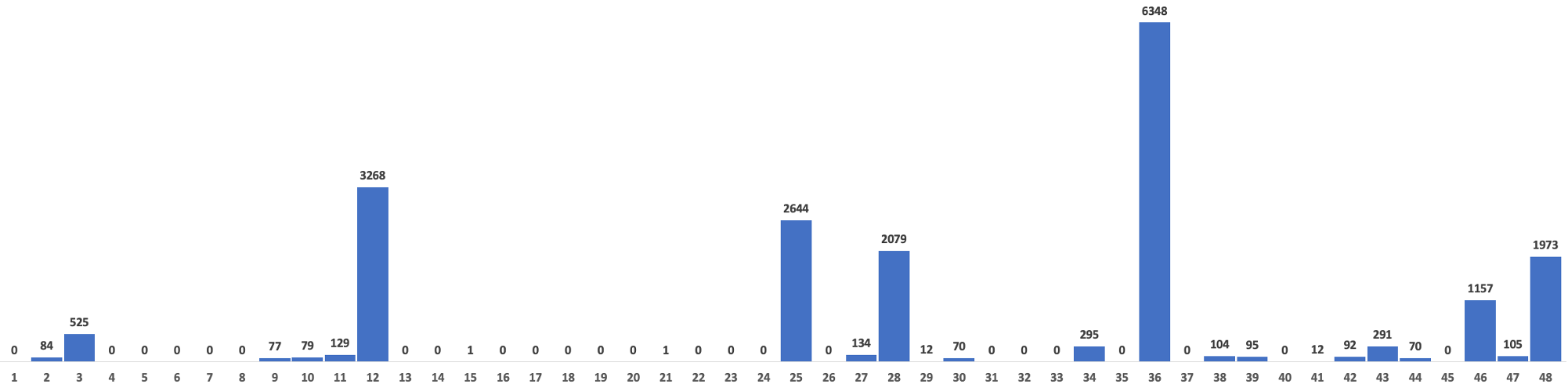}
    \caption{Depiction of the number of videos the $48$ users created in  over $8$ weeks.}
    \label{fig:create}
\end{figure*}

\begin{figure*}
    \includegraphics[width=\linewidth]{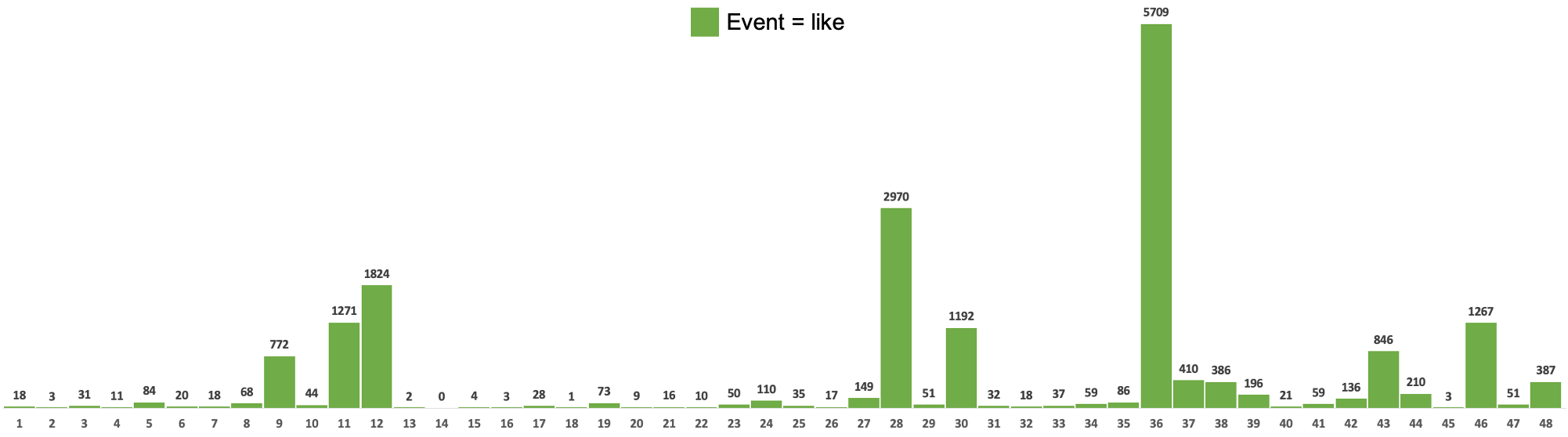}
    \caption{Depiction of the number of videos the $48$ users liked in  over $8$ weeks.}
    \label{fig:like}
\end{figure*}

\begin{figure*}
    \includegraphics[width=\linewidth]{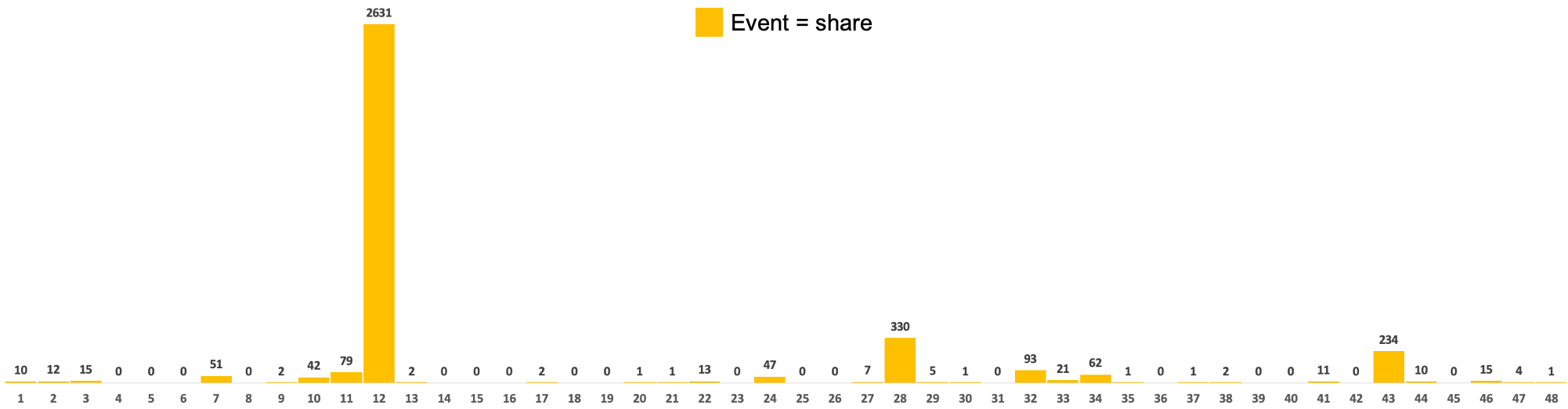}
    \caption{Depiction of the number of videos the $48$ users shared in  over $8$ weeks.}
    \label{fig:share}
\end{figure*}

\begin{figure*}
    \includegraphics[width=\linewidth]{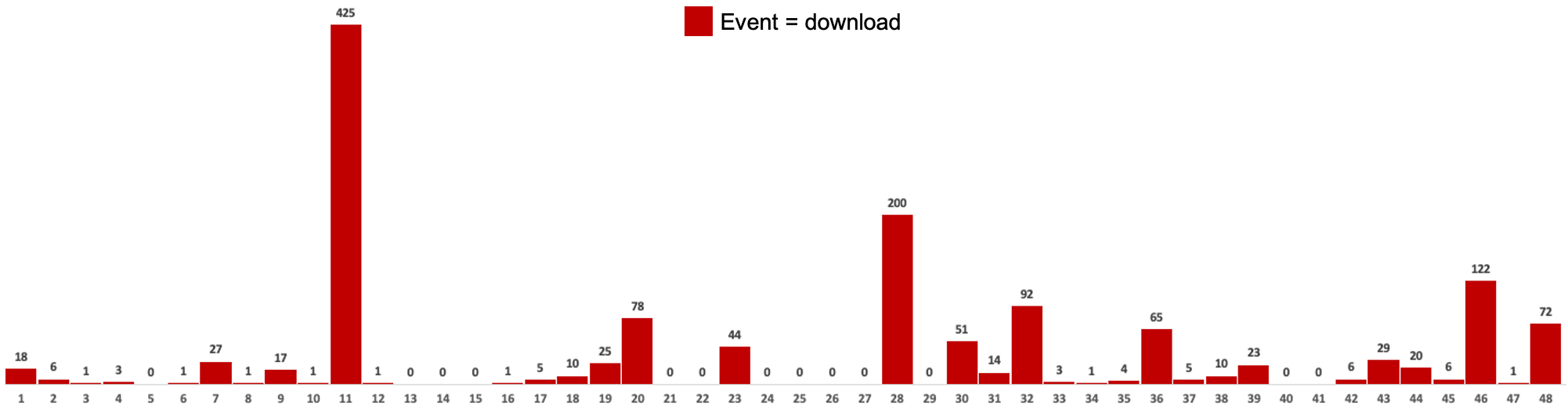}
    \caption{Depiction of the number of videos the $48$ users downloaded in  over $8$ weeks.}
    \label{fig:download}
\end{figure*}

\begin{figure*}
    \includegraphics[width=\linewidth]{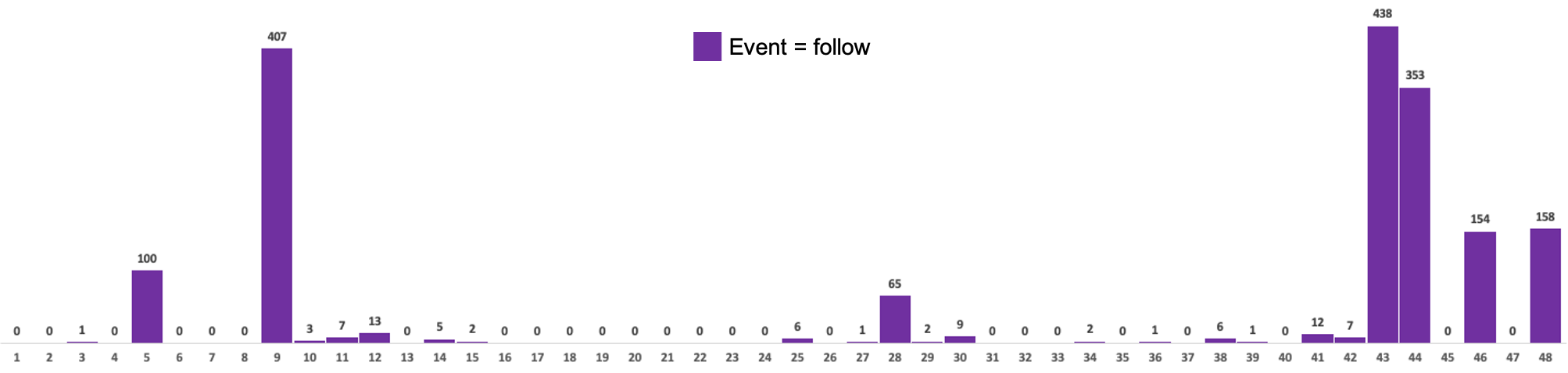}
    \caption{Depiction of the number of videos the $48$ users followed in  over $8$ weeks.}
    \label{fig:follow}
\end{figure*}

\begin{figure*}
    \includegraphics[width=\linewidth]{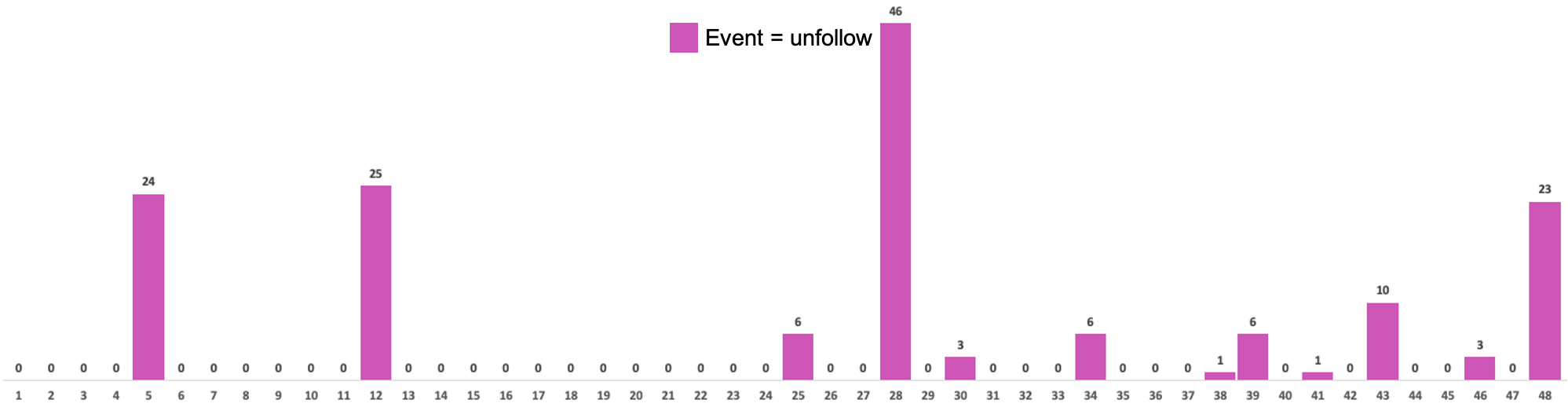}
    \caption{Depiction of the number of videos the $48$ users unfollowed in  over $8$ weeks.}
    \label{fig:unfollow}
\end{figure*}

Figure~\ref{fig:neighbors} lists the neighbor count for each central user; the highest being $6,587$. A neighbor is any user whose content has been engaged with by a central user, and who may not necessarily be a follower of, or follow, the central user. Similarly, in Figure~\ref{fig:topic}, we visually depict the breadth of topics the $48$ users engaged in over the entire time span. Finally, in Figure~\ref{fig:play} - Figure~\ref{fig:unfollow}, we depict the frequency of each action in $\mathcal{A}$ by all $48$ users. 

\newpage
\section{More Experiments and Results}
\label{sec:more-expts}
We first present more case studies for more analysis of Q$1-4$ in Section~\ref{subsec:case-study}. Then in Section~\ref{subsec: efficiency_analysis}, we discuss the efficiency of our approach in terms of computational resources highlighting the benefits of modeling localized emotion contagion in dynamic social network graphs. 
\subsection{More User Study Analysis}
\label{subsec:case-study}
We add $2$ more case studies to extend our evaluation in Section $6.2.1$ for Q$1-4$.

\noindent \textbf{Case Study $\bm{4}$:} \textit{User A watches a mix of "Romantic Poetry" videos for a month, often downloading and watching videos by users D and E. User A creates a video in this category after 1 month. User A does not normally create many videos in this category.}

The focus of this study is to confirm the contagion resulting from having created a video after watching many similar videos in the same category. This time, we ask participants if both D and E were responsible for causing contagion. An overwhelming majority replied affirmatively, which is also corroborated by our model ($\xi_D = 11.86, \xi_E = 3.27$).

\noindent \textbf{Case Study $\bm{5}$:} \textit{User A watched 1 video posted by user C and unfollowed user C afterwards. During the next 3 weeks, user A watched a few more videos posted by user C and followed user C again.}

The focus of this study is to confirm the contagion resulting from having followed a user after watching their videos. As before we ask participants if C is responsible for causing contagion. An overwhelming majority replied affirmatively, which is also corroborated by our model ($\xi_C = 2.53$). In addition, we also note that the profile embedding similarity between the central user A and user C is $0.63$.
\subsection{Computation efficiency Analysis}
\label{subsec: efficiency_analysis}
Global diffusion involves simultaneously computing the contagion effects for every central user for every neighbor \textit{simultaneously} by diffusing through the entire graph~\cite{fan2018agent}. But the cost of storing the audio-visual information for the entire graph scales with the number of central users and the average number of neighbors per user. Empirically, we found that for a graph consisting of $50$ central users, each with $1,461$ neighbors on average, modeling the global diffusion would require approximately $4.22$ days and $14.60$ TB of storage. Isolating the computation across central users and neighbors, on the other hand, reduces the cost by a factor of $\bigO{CM}$, where $C$ and $M$ denote the number of central users and average number of neighbors per user, respectively. In our setup, computing the emotion contagion requires approximately $2$GB and $5$ minutes.

Computing contagion locally offers several benefits in addition to reducing the computational complexity. By performing local diffusion over selected users, we avoid bottlenecks in global diffusion caused by including popular or famous users who, though few, typically contain thousands or even millions of followers (neighbors). Local diffusion also elegantly handles addition and deletion of edges in realtime; a change in the edge list for a central user $c$ does not affect local diffusion for all remaining central users. This is not true in the case of global diffusion, where changes in the edge structure in any part of the graph would necessitate restarting the diffusion process.

\newpage
\section{Ethical Considerations}
\label{sec:ethical-considerations}
The dataset used in this paper sources videos from a popular social media application for sharing short videos. These videos show users' faces and their user profiling information contains personal details such as age, gender, language, and location. 
Given the sensitive nature of this dataset, we decide \textit{against} publicly releasing the data. We have, instead, provided directions to replicate our data preparation process on other social media platforms. More importantly, we do \textit{not} collect \textit{any} personal information of the involved human participants in the user studies. 
\end{document}